# The episode of Quasi-Biennial Oscillation in 2016 did not have its roots in its distant past; the role of the strong El Niño event

Costas A. Varotsos (1) Nikos V. Sarlis (2) and Maria Efstathiou (1)

*Climate Research Group, Section of Environmental Physics and Meteorology, Dept. of Physics, National and Kapodistrian University of Athens, Campus Bldg. Phys. V, Athens 15784, GR (1), Section of Solid State Physics, Dept. of Physics, National and Kapodistrian University of Athens, Panepistimiopolis Zografos, Athens 157 84, GR (2)*

**It is generally accepted that the Earth's climate system is a subsystem of a wider system of the so-called global system and displays various modes of variability[1-8]. The spatio - temporal features of the main modes of natural variability have not yet been accurately simulated, while an unknown number of low-frequency modes have not yet been identified, mainly due to comparatively short observational data records and climate noise components[9-12]. The available advanced models are capable of accurately simulating one or more modes of climate variability, but there is not a single model (or an array of them) that performs reliable simulations across all modes of variability. For example, today, less than a third of the general circulation models can reliably reproduce the first three most dominant and very regular quasi-periodical oscillations of the climate system. A classic example of quasi-periodic oscillation is the Quasi-Biennial Oscillation (QBO), which refers to a normal cycle of 2 to 3 years in several atmospheric quantities such as zonal wind and ozone[13-20]. We are investigating the abnormal behavior of the equatorial QBO in the zonal wind that exhibits it, since February 2016 [21]. More specifically, QBO broke off the normal pattern and the eastward stratospheric winds unexpectedly reversed to a westward direction. We focus on the temporal evolution of the equatorial zonal wind in the altitude region 70 - 10 hPa, investigating in particular whether this unprecedented event could be seen as a result of scaling effect in the equatorial zonal wind.**

Along these lines we use the monthly mean values of the zonal wind over Singapore, at the pressure levels 10, 20, 30, 40, 50, 70 hPa [22] during the period 1987-2016, as well as the monthly mean values of zonal wind over the equator at 30 hPa, during 1948-2016, as computed from the NCEP / NCAR Reanalysis. We also use the monthly mean anomalies time series of the Niño 1+2 and Niño 3 SST indices as three month running means. The analysis technique used in order to study the intrinsic properties of zonal wind time series was the detrended fluctuation analysis (DFA), which eliminates the noise of the nonstationarities that characterize the zonal wind and permits the detection of intrinsic self-similarity[23-26]. Moreover, to further investigate the temporal evolution of the sensitivity of the wind field, both time series (1948-2016 and 1987-2016) of the monthly mean values of the zonal wind at 30 hPa have been also analyzed during their overlapping period by using a new time domain termed natural time which uncovers hidden dynamic properties [27]

**Has the cause of the QBO disruption in 2016 originated from the past?**

It is noteworthy that the exploration of the temporal evolution of the El Niño Southern Oscillation - ENSO from Jan 1876 to Nov 2011 by means of natural time revealed that the major ENSO extremes provide precursor signals that are maximized in a time window of 2 years on average, i.e. a QBO cycle [28].

It should be emphasized that in the 3 to 4 years after the warm ENSO events (i.e. 1982, 1997, and 2015), the QBO and the ENSO were aligned indicating that strong warm ENSO events can lock the phase of the QBO[29]. However, the evolution of the latest warm ENSO needs further investigation. We stress the point that it was unusual owing not only to its NH summer onset, but the way in which the tropical Pacific atmosphere was affected, displaying unusual features, which may have affected the QBO [30, 31].

It is very important to investigate whether the equatorial zonal wind in the region between the pressure levels 10-70hPa obeys scaling behavior. For this purpose we have used the monthly mean zonal wind components (0.1 m/s) at Singapore provided by the Free University of Berlin (http://www.geo.fu-berlin.de/en/index.html). The results obtained from the detrended fluctuation analysis (DFA) revealed persistence with a crossover at roughly 25 months that resulted from the QBO signal. To establish however the power-law scaling in the equatorial zonal wind two criteria should be fulfilled: the constancy of local slopes in a sufficient range and the rejection of the exponential decay of the autocorrelation function must be confirmed. While the local slopes pattern shows that they reach a constant level at long scales, the data do not meet the second criterion, i.e. rejection of the exponential decay of the autocorrelation function.

Then the same analysis was repeated taking into account the QBO time series (30hPa zonal wind at the equator) calculated at NOAA/ESRL Physical Science Division for the period 1948- 2016. The results showed that the local slopes reach a constant value at long scales, but the rejection of the exponential decay of the autocorrelation function fails. Moreover, we reach the same conclusion, when using the QBO time series of 30 hPa zonal wind index during 1979-2016, obtained from NOAA/ Climate Prediction Center.

**The possible link between the last strong ENSO and QBO disruption events**

We now turn our attention to the results obtained from the natural time analysis. It interestingly shows for both data sets (related with 30hPa) a precursory behavior before the maximization of the zonal wind velocity and that the recent strong El Niño event might be related with the unprecedented QBO behavior. In conclusion the data of the equatorial zonal wind field do not satisfy both the above-mentioned criteria and thus the long-range correlations cannot be established. In other words, the unprecedented event of the equatorial QBO in zonal wind this year could not be considered as a result of scaling, viz. that has not its roots in the distant past. However, it should be noted that this unexpected QBO event was not anticipated by forecast models and hopefully it will motivate a better understanding of driving mechanisms of QBO (cf. natural time analysis may provide precursory phenomena before the maximization of the zonal wind speed).

**Note:** A preliminary version of this work was presented at the first CVAS workshop "What do we know about multicentennial, multimillenial variability? 28-30 November 2016, Hamburg, Germany.

**Competing Interests** The authors declare that they have no competing financial interests.

**Correspondence** Correspondence and requests should be addressed C. Varotsos (email: covar@phys.uoa.gr)